\documentclass{svmult}

\usepackage{makeidx}         
\usepackage{graphicx}        
\usepackage{multicol}        
\usepackage[bottom]{footmisc}
\usepackage{epsfig}
\makeindex             

\begin{document}
\title*{Networks of companies and branches in Poland}
\author{A. M. Chmiel \and
J. Sienkiewicz \and K. Suchecki \and J. A. Ho{\l}yst} \institute{Faculty of Physics and Center of Excellence for
Complex Systems Research\\ Warsaw University of Technology, Koszykowa 75, PL 00-662 Warsaw, Poland
\texttt{jholyst@if.pw.edu.pl} }

\maketitle
\section{Introduction}
During the last few years various models of networks \cite{przegl1,przegl2} have become a powerful tool for
analysis of complex systems in such distant fields as Internet \cite{net1}, biology \cite{bio1}, social groups
\cite{social1}, ecology \cite{eco1} and public transport \cite{julian1}. Modeling behavior of economical agents is
a challenging issue that has also been studied from a network point of view. The examples of such studies are models of
financial networks \cite{CBG}, supply chains \cite{HL,HL2}, production networks \cite{WB}, investment networks
\cite{BZ} or collective bank bankrupcies \cite{agata1,agata2}. Relations between different companies have been
already analyzed using several methods: as networks of shareholders \cite{GSB}, networks of correlations between
stock prices \cite{stock1} or networks of board directors \cite{board}. In several cases scaling laws  for network
characteristics have been observed.\\

In the present study we consider relations between companies in Poland taking into account common branches they
belong to. It is clear that companies belonging to the same branch compete for similar customers, so the market
induces correlations between them. On the other hand two branches can be related by companies acting in both of
them. To remove weak, accidental links we shall use a concept of threshold filtering for weighted networks where a
link weight corresponds to a number of existing connections (common companies or branches) between a pair of
nodes.

\section{Bipartite graph of companies and trades}

\begin{figure}[ht]
\vskip 1cm
 \centerline{\epsfig{file=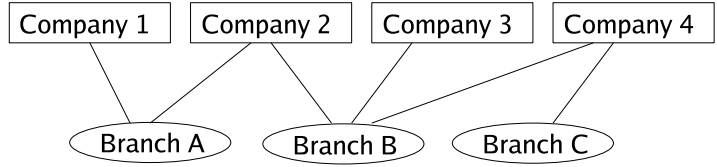,width=.8\textwidth}}
    \caption{Bipartite graph of companies and trades.}
    \label{fig:bi}
\end{figure}

We have used the commercial database "Baza Kompass Polskie Firmy B2B" from September 2005. It contains information
about over 50 000 large and medium size Polish companies belonging to one or more of 2150 different branches. We
have constructed a bipartite graph of companies and trades in Poland as at Fig. \ref{fig:bi}.

In the bipartite graph we have two kinds of objects: branches $A=1,2,3....N_b$ and companies $i=1,2,3......N_f$,
where $N_b$ -- total number of  branches and  $N_f$ -- total number  of companies. Let us define a {\em branch
capacity} $|Z(A)|$ as the cardinality of set of companies belonging to the branch $A$. At Fig. \ref{fig:bi} the
branch $A$ has the capacity $|Z(A)|=2$ while $|Z(B)|=3$ and $|Z(C)|=1$. The largest capacity of a branch in our
database was $2486$ (construction executives), the second largest was $2334$ (building materials).

Let $B(i)$ be a set of branches a given company $i$ belongs to. We define a {\em company diversity} as $|B(i)|$.
An average company diversity $\mu$ is given as

 \begin{equation}\label{eq:mu}
    \mu=\frac{1}{N_f} \sum^{i=N_f}_{i=1}|B(i)|
\end{equation}
For our data set we have $\mu=5.99$.\\ Similarly an average branch capacity $\nu$ is given as
\begin{equation}\label{eq:mu2}
    \nu=\frac{1}{N_b}\sum^{A=N_b}_{A=1}|Z(A)|
\end{equation}
and we have $\nu=134$.\\ It is obvious that the following relation is fulfilled for our bipartite graph:
\begin{equation}\label{eq:n}
    \frac{\nu}{N_f}=\frac{\mu}{N_b}.
\end{equation}

\section {Companies and trades networks}
\begin{figure}[ht]
\vskip 1cm
 \centerline{\epsfig{file=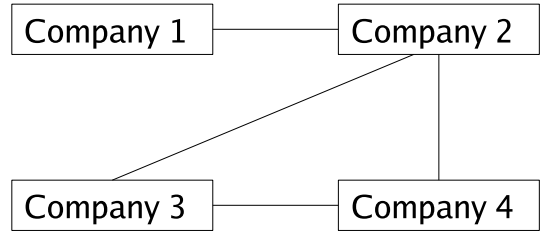,width=.6\textwidth}}
    \caption{Companies network.}
    \label{fig:bf}
\end{figure}

The bipartite graph from Fig. \ref{fig:bi} has been transformed to create a {\em companies network}, where nodes
are companies and a link means that two connected companies belong to at least one common branch. If we used the
example from Fig.\ref{fig:bi} we would obtain a companies network presented at Fig. \ref{fig:bf}.

We have excluded from our dataset all items that correspond to communities (local administration) and for our
analysis we consider $N_f=48158$ companies. All companies belong to a single cluster. Similarly a {\em trade
(branch) network} has been constructed where nodes are trades and an edge represents connection if at least one
company belongs to both branches. In our database we have $N_b=2150$ different branches.
\begin{figure}[ht]
\vskip 1cm
 \centerline{\epsfig{file=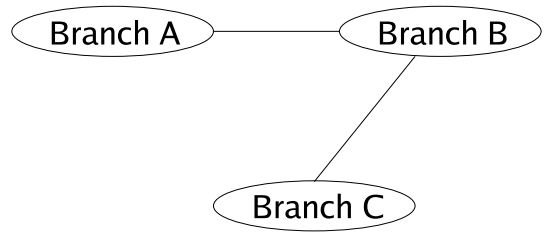,width=.6\textwidth}}
    \caption{Trades network.}
    \label{fig:br}
\end{figure}

\section {Weight, weight distribution and networks with cutoffs}
\begin{figure}[ht]
\vskip 1cm
 \centerline{\epsfig{file=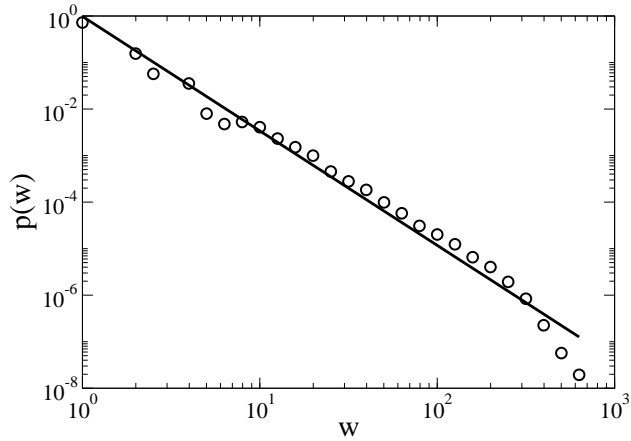,width=.7\textwidth}}
    \caption{Weight distribution in branches network. }
    \label{fig:wb1}
\end{figure}

We have considered link-weighted networks. In the branches network the link weight means a number of companies that are active  in the same pair  of branches 
and it is formally a cardinality of a common part of sets $Z(A)$ and $Z(B)$, where $Z(A)$ is a set of companies
belonging to the branch $A$ and $Z(B)$ is a set of companies belonging to the branch $B$.

\begin{equation}\label{eq:w}
    w_{AB}= |Z(A) \cap Z(B)|
\end{equation}
Let us define a function $f_k^A$ which is equal to one if a company $k$ belongs to the branch $A$, otherwise it is
zero.

\begin{equation}\label{eq:w1}
    f_k^A=\left\{\begin{array}{l}1 ,k \in A\\0,k\notin A\end{array}\right\}
\end{equation}
Using the function $f_k^A$ the weight can be written as:
\begin{equation}\label{eq:w2}
 w_{AB}=\sum_{k=1}^{N_F} f_k^A f_k^B
\end{equation}
The weight distribution $p(w)$, meaning the probability $p$ to find a link with a given weight $w$, is presented
at Figure \ref{fig:wb1}. The distribution is well approximated by a power function
\begin{equation}\label{eq:pw}
  p(w)\sim w^{-\gamma}
\end{equation}

where the exponent $\gamma=2.46\pm 0.07$. One can notice the existence of edges with large weights. The maximum
weight value is $w_{max}=764$, and the average weight
\begin{equation}\label{eq:ws}
    \langle w\rangle =\sum^{w_{max}}_{w_{min}}wp(w)
\end{equation}
equals $\langle w\rangle=4.67$.\\

Using cutoffs for link weights we have constructed networks with different levels of filtering. In such networks
nodes are connected only when their edge weight is no less than an assumed cutoff parameter $w_o$.

\begin{table}
\centering \caption{Data for branches networks: $w_o$ is the value of selected weight cutoff, $N$ is the number of
vertex with nonzero degrees, $E$ is the number of links, $k_{max}$ is the maximum node degree, $\langle k\rangle$
is the average node degree, $C$ is the clustering coefficient.} \label{tab:branze}

\begin{tabular}{l@{\hspace{0.5cm}}l@{\hspace{0.5cm}}@{\hspace{0.5cm}}l@{\hspace{0.5cm}}l@{\hspace{0.5cm}}l@{\hspace{0.5cm}}l@{\hspace{0.5cm}}l}
\hline\noalign{\smallskip}
  $w_o$ & $N$ & $E$ &$k_{max}$  & $\langle k\rangle$ & $C$  \\
\noalign{\smallskip}\hline\noalign{\smallskip}

1   &   2150    &   389542  &   1716    &   362 &   0.530   \\ 2   &   2109    &   212055  &   1381    & 201 &
0.565   \\ 3   &   2053    &   136036 & 1127    &   132 &   0.568   \\ 4   & 2007    & 100917  &   952 &   100 &
0.575   \\ 5   &   1948    &   80358   &   802 &   82  &   0.589   \\ 1   &   2150    &   389542  & 1716 &   362 &
0.530   \\ 2   &   2109    &   212055  &   1381    &   201 &   0.565   \\ 3 &   2053    & 136036  & 1127    & 132.
&   0.568   \\ 4   &   2007    &   100917  &   952 & 100 &   0.575   \\ 5   &   1948 & 80358   &   802 & 82  &
0.589   \\ 6   &   1904 &   66353   &   655 &   69  & 0.592   \\ 7   &   1858    & 56565   & 569 &   60 &   0.596
\\ 8   & 1819 &   49193 & 519 &   54 &   0.597 \\ 9   & 1786    & 43469   & 477 &   48  &
0.599 \\ 10  & 1748 &   38924 &   450 & 44  &   0.600   \\ 12 & 1666    &   32167   & 394 & 38  &   0.615   \\ 14
&   1611    &   26088   &   325 &   32  & 0.605 \\ 16  & 1545    & 21762   &   288 & 28  & 0.606   \\
18  &   1490    & 18451   & 259 &   24  &   0.603 \\ 20  &   1424    &   15872   &   226 &   22  &   0.604   \\ 30
& 1188    &   8989    & 162 &   15  & 0.585   \\ 40  &   996 &   6036    &   131 &   12 &   0.587 \\ 50 & 857 &
4379    & 111 &   10  &   0.572   \\ 60  &   752 &   3303    & 85 &   8   & 0.551   \\ 70  &   666 &   2638    &
65  &   7   &   0.524   \\ 80  &   575 & 2143    &   55 &   7 &   0.532 \\ 90  &   512 &   1808    &   49  & 7   &
0.538   \\ 100 &   464 &   1543    &   41  &   6   &   0.546   \\ 150 &   306 & 750 &   26  &   4   & 0.493 \\
\noalign{\smallskip}\hline
\end{tabular}
\end{table}

A weight in the companies network is defined in a similar way as in the branches networks, i.e. it is the number
of common branches for two companies --- formally it is equal to the cardinality of a common part of sets $B(i)$
and $B(j)$, where $B(i)$ is a set of branches the company $i$ belongs to, $B(j)$ is a set of branches the company
$j$ belongs to.

\begin{equation}\label{eq:w3}
    w_{ij}= |B(i) \cap B(j)|
\end{equation}
Using the function $f_k^A$ the weight can be written as
\begin{equation}\label{eq:w4}
 w_{ij}=\sum_{A=1}^{N_b}f_i^A f_j^A.
\end{equation}

The maximum value of observed weights $w_{max}=207$ is smaller in this networks than in the branches network while
the average value equals $\langle w\rangle=1.48$. The weight distribution is not a power law in this case and it
shows an exponential behavior in a certain range.

\begin{figure}[ht]
\vskip 1cm
\centerline{\begin{tabular}{ c c}
 \epsfig{file=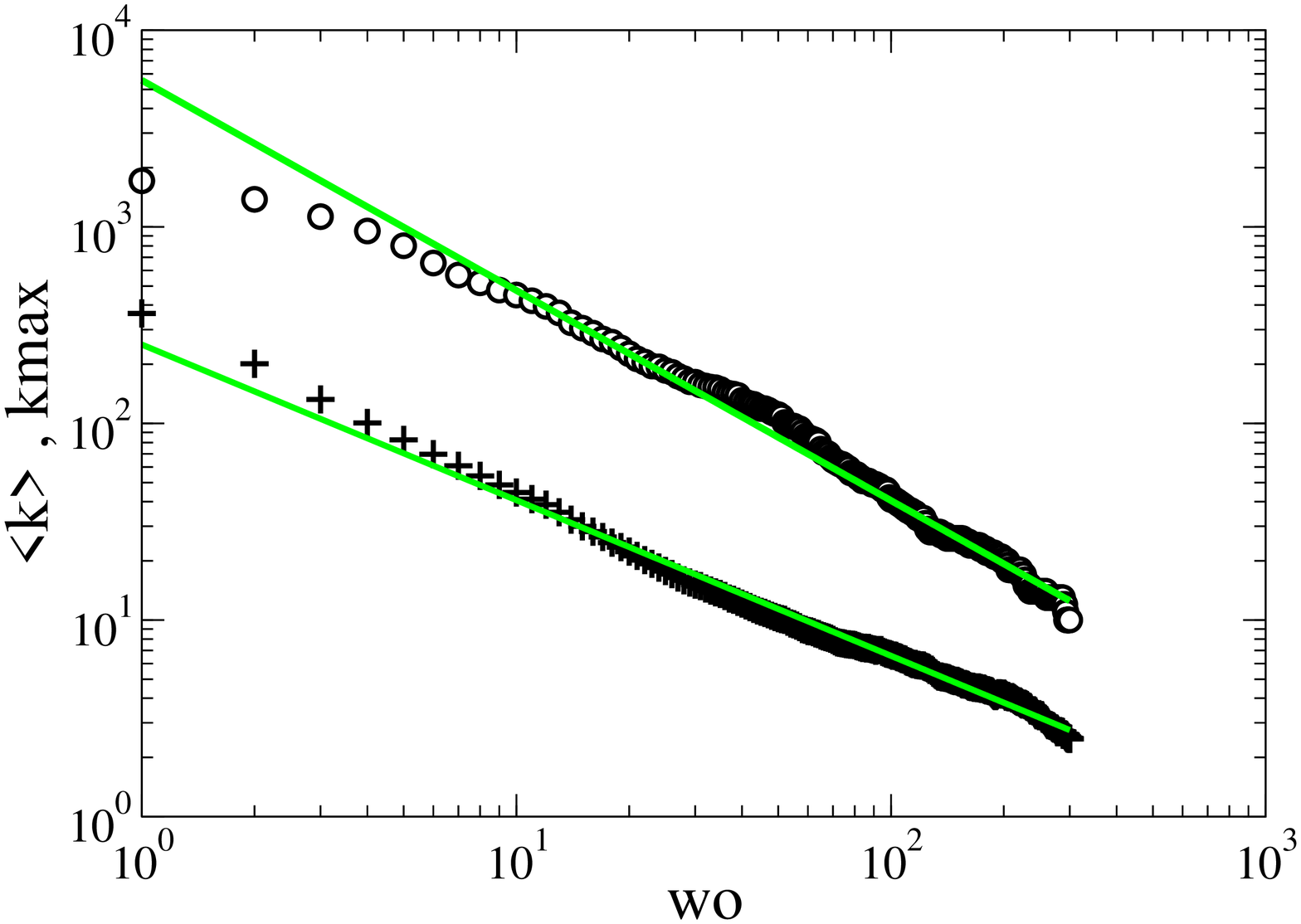,width=.5\textwidth} & \epsfig{file=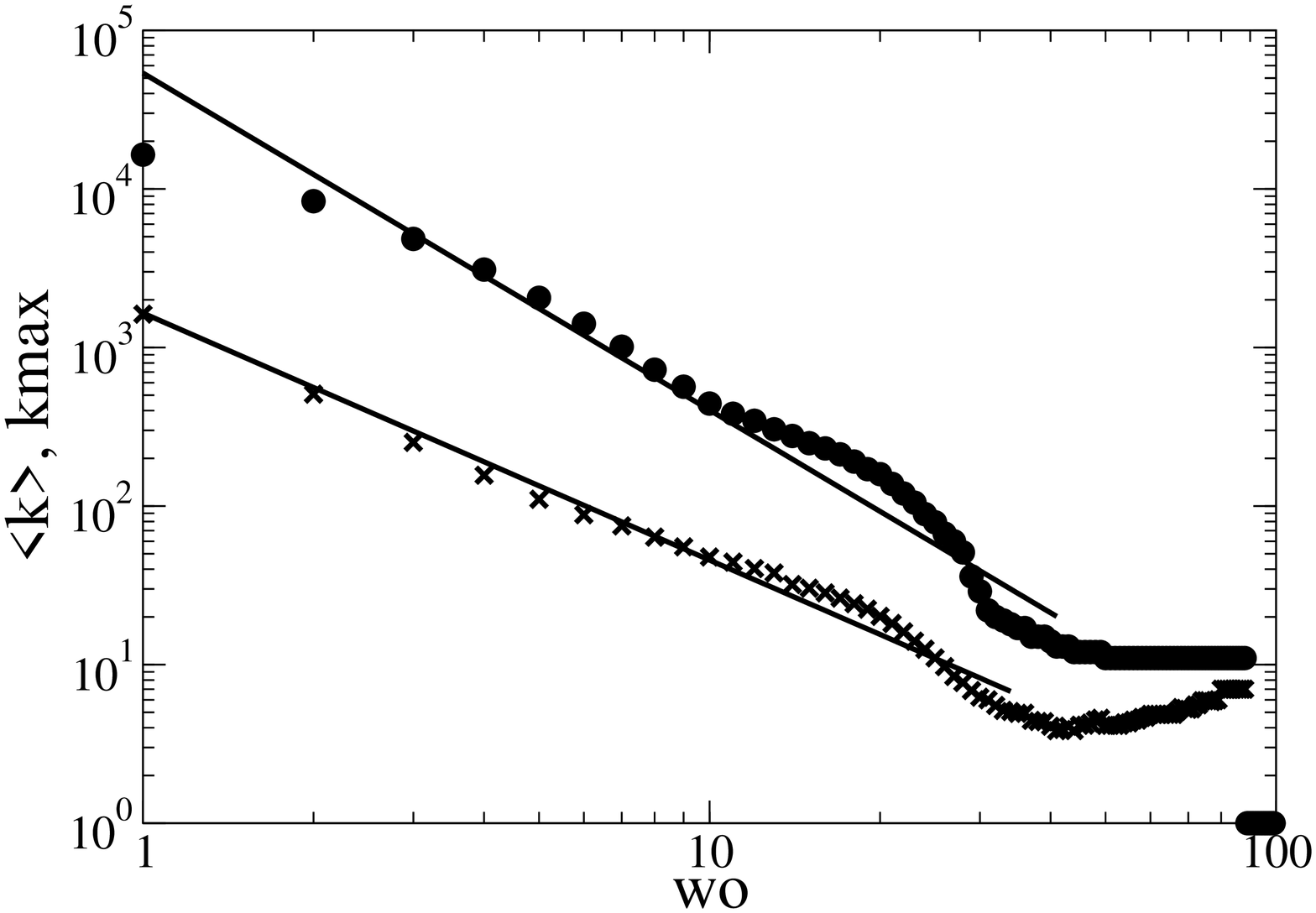,width=.5\textwidth} \\
\end{tabular}}
    \caption{Dependence of $\langle k\rangle$ and $k_{max}$ on cutoff parameter $w_o$ for branches networks (left) and companies networks (right).}
    \label{fig:kkmax}
\end{figure}

Similarly to the branches networks we have introduced cutoffs in companies network. At the Fig.\ref{fig:kkmax} we
present average degrees of nodes and maximum degrees as functions of the cutoff parameter $w_o$.
We have observed a power law scaling
\begin{equation}\label{eq:kkmax}
    \langle k\rangle \sim w_o^{-\beta}
\end{equation}
\begin{equation}\label{eq:kkmax1}
    k_{max} \sim w_o^{-\alpha}
\end{equation}

where for branches networks $\alpha_b=1.069\pm 0.008$ and $\beta_b= 0.792 \pm 0.005$ while for companies networks
$\alpha_f=2.13 \pm 0.07$ and $ \beta_f=1.55\pm 0.04$.

\begin{table}
\centering \caption{Data for companies networks: $w_o$ is the selected cutoff, $N$ is the number of nodes with
nonzero degrees, $E$ is the number of links, $k_{max}$ is the maximum node degree, $\langle k\rangle$ is the
average node degree, $C$ is the clustering coefficient.} \label{tab:f}

\begin{tabular}{l@{\hspace{0.5cm}}l@{\hspace{0.5cm}}l@{\hspace{0.5cm}}l@{\hspace{0.5cm}}l@{\hspace{0.5cm}}l@{\hspace{0.5cm}}l@{\hspace{0.5cm}}l}
\hline\noalign{\smallskip}
  $w_o$ & $N$ & $E$ &$k_{max}$  & $\langle k\rangle$ & $C$ \\
\noalign{\smallskip}\hline\noalign{\smallskip}

1   &   48158   &   39073685    &   16448   &   1622    &   0.652   \\
2   &   39077   &   9932790 &   8366    &   508 &   0.689   \\
3   &   31150   &   3928954 &   4842    &   252 &   0.714   \\
4   &   24212   &   1895373 &   3103    &   156 &   0.717   \\
5   &   18566   &   1024448 &   2059    &   110 &   0.713   \\
6   &   14116   &   622662  &   1412    &   88  &   0.710   \\
7   &   10796   &   404844  &   1012    &   74  &   0.700   \\
8   &   8347    &   266013  &   724 &   63  &   0.701   \\
9   &   6527    &   180696  &   566 &   55  &   0.699   \\
10  &   5197    &   124079  &   443 &   47  &   0.699   \\
11  &   4268    &   94531   &   382 &   44  &   0.704   \\
12  &   3400    &   68648   &   345 &   40  &   0.693   \\
13  &   2866    &   54258   &   305 &   37  &   0.691   \\
14  &   2277    &   36461   &   277 &   32  &   0.663   \\
15  &   1903    &   28844   &   249 &   30  &   0.673   \\
16  &   1627    &   23063   &   231 &   28  &   0.678   \\
17  &   1397    &   18352   &   212 &   26  &   0.667   \\
18  &   1196    &   14480   &   191 &   24  &   0.680   \\
19  &   1003    &   11230   &   171 &   22  &   0.680   \\
20  &   883 &   8907    &   159 &   20  &   0.676   \\
\noalign{\smallskip}\hline
  \end{tabular}
\end{table}

\section{Degree distribution}
\begin{figure}[ht]
\vskip 1cm
 \centerline{\epsfig{file=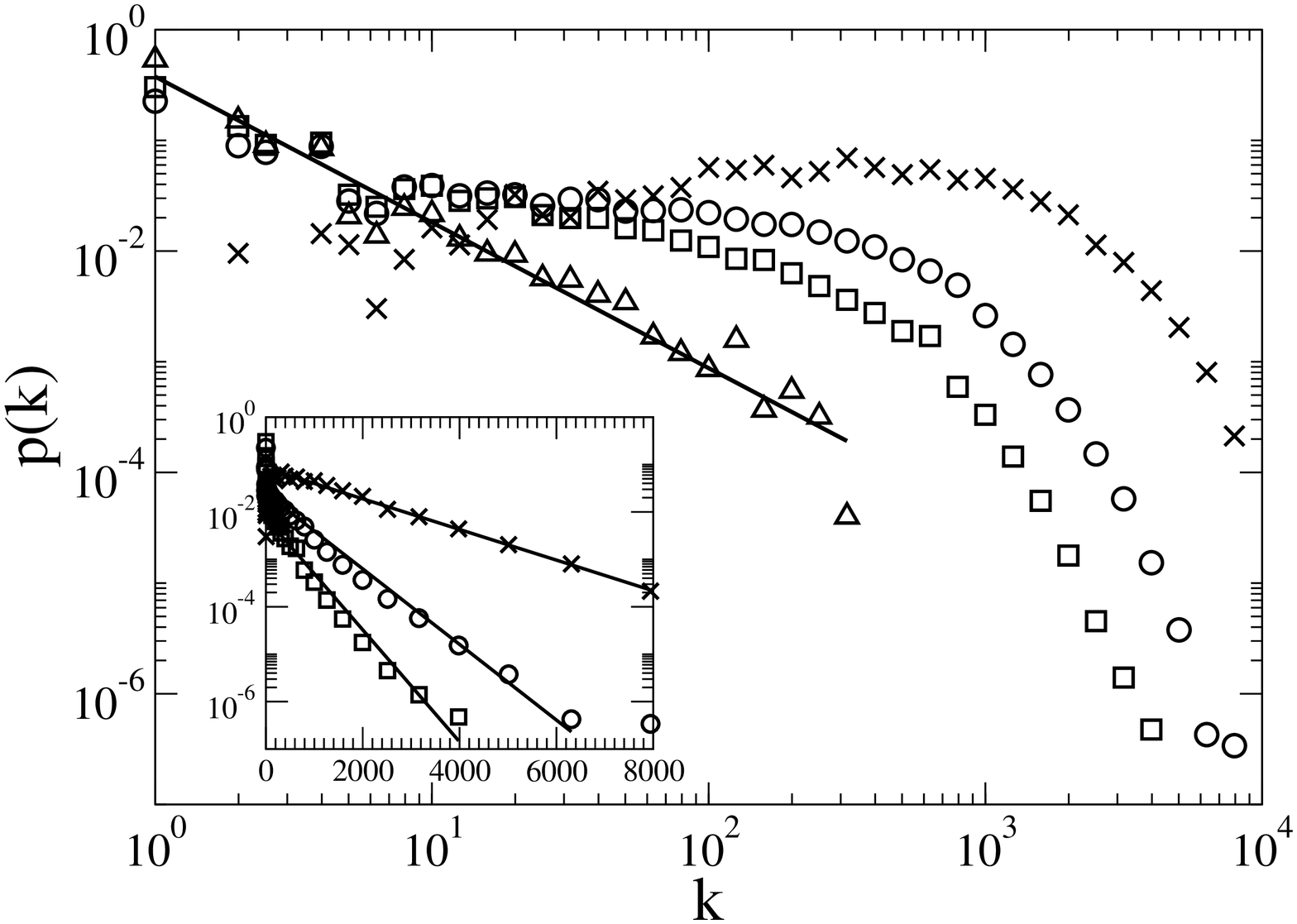,width=.8\textwidth}}

    \caption{Degree distributions for companies networks for different values of $w_o$. X-marks are for $w_o=1$, circles are for $w_o=2$,
    squares are for $w_o=3$ and triangles are for $w_o=12$.}
    \label{fig:pf0}
\end{figure}

We have analyzed the degree distribution for networks with different cutoff parameters. At Fig. \ref{fig:pf0} we
present the degree distributions for companies networks for different values of $w_o$. The distributions change
qualitatively with increasing $w_o$ from a nonmonotonic function with an exponential tail (for $w_o=1$) to a power
law with exponent $\gamma$ (for $w_o>6$).

Values of exponent $\gamma$ for different cutoffs are given in the Table \ref{tab:fr}.
\begin{table}
\centering

\caption{Values of exponent $\gamma$ for different cutoffs $w_o$ in companies networks.} \label{tab:fr}

\begin{tabular}{l@{\hspace{0.5cm}}l@{\hspace{0.5cm}}l}
\hline\noalign{\smallskip}

 $w_o$ & $\gamma$ & $\Delta \gamma$  \\
\noalign{\smallskip}\hline\noalign{\smallskip}
6 & 1.06  &0.03\\
8  &1.12  &0.04\\
10 &1.22 & 0.05\\
12 &1.23 & 0.06\\
14& 1.31 & 0.05\\
16 &1.31 & 0.06\\
18 &1.37  &0.07\\
20 &1.35 & 0.07\\
\noalign{\smallskip}\hline
\end{tabular}
\end{table}

\begin{figure}[ht]
\vskip 1cm
\centerline{\epsfig{file=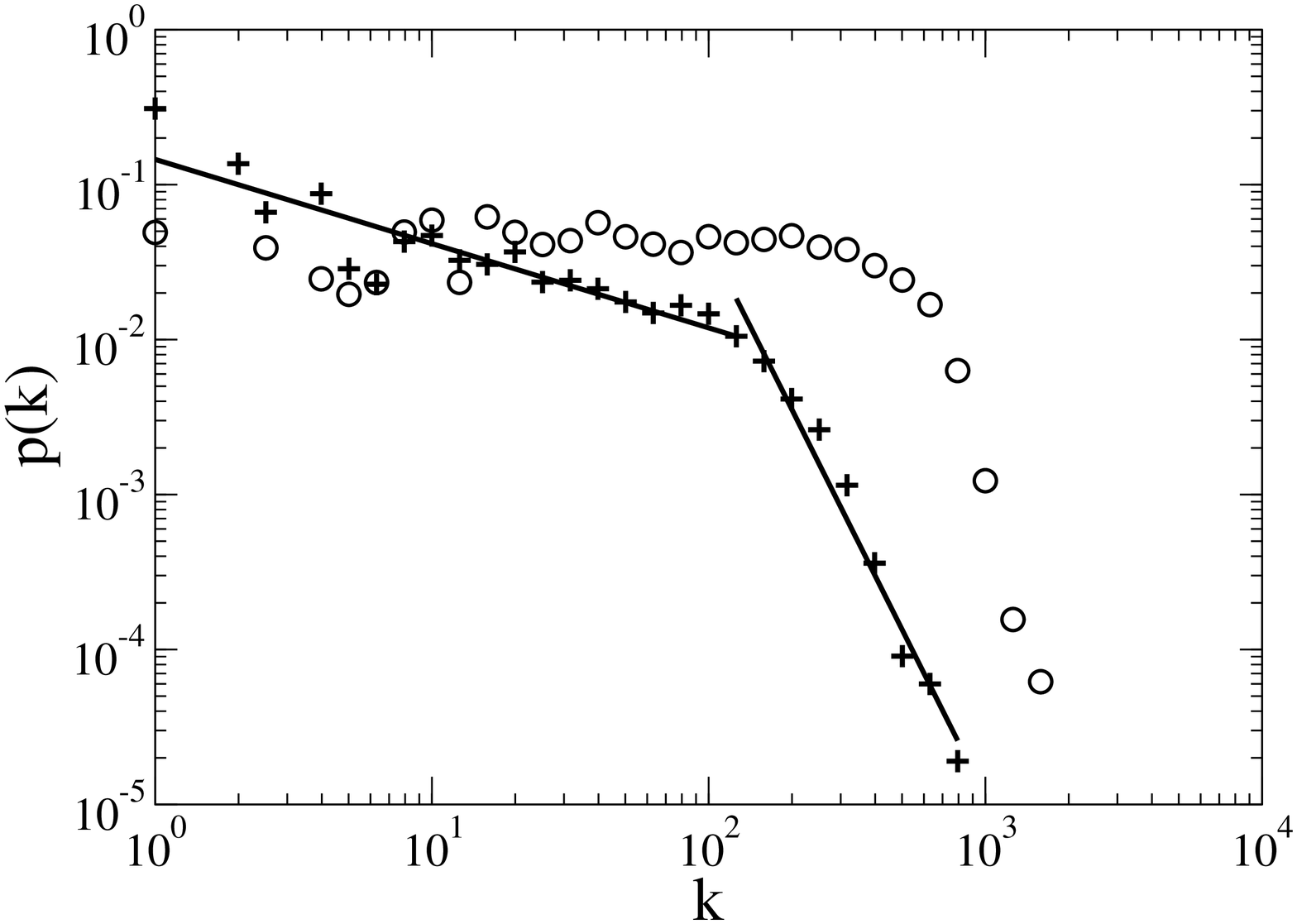,width=.8\textwidth} }
    \caption{Degree distribution in branches network for different values of $w_o$. Circles are for $w_o=1$, crosses are for $w_o=4$.}
    \label{fig:pw0a}
\end{figure}

Now let us come back to branches networks. At the Fig. \ref{fig:pw0a} we present a degree distribution for
$w_o=1$. We observe a high diversity of node degrees --- vertices with large values of $k$ occur almost as
frequent as vertices with a small $k$.

For a properly chosen cutoff values the degree distributions are described by power laws. For $w_o=4$ we see two
regions of scaling with different exponents $\gamma_1$ and $\gamma_2$ while a transition point between both
scaling regimes appears at $k\approx 100$. The transition appears due to the fact that there are almost no
companies with diversity over $100$, so branches with $k>100$ have connections due to several companies, as
opposed to branches with $k<100$ that can be connected due to a single company. However the probability that many
companies link a single branch with many different others is low, thus the degree probability $p(k)$ decays much
faster after the transition point. In the Table \ref{tab:wspol} we present values $\gamma_1$ and $\gamma_2$ for
different cutoffs $w_o$.\\

\begin{table}
\centering \caption{Values of scaling exponents $\gamma_1$ and $\gamma_2$ for branches networks.}\label{tab:wspol}
  \begin{tabular}{l@{\hspace{0.5cm}}l@{\hspace{0.5cm}}l@{\hspace{0.5cm}}l@{\hspace{0.5cm}}l}
  \hline\noalign{\smallskip}
        $w_o$ & $\gamma_1$ & $\Delta \gamma_1$ &  $\gamma_2$& $\Delta \gamma_2$  \\
\noalign{\smallskip}\hline\noalign{\smallskip}

4   &       0.54    &   0.06    &   3.56    &   0.22    \\
5   &       0.59    &   0.05    &   3.70    &   0.21    \\
6   &       0.62    &   0.06    &   3.60    &   0.22    \\
7   &       0.64    &   0.07    &   3.44    &   0.19    \\
8   &       0.69    &   0.06    &   3.53    &   0.22    \\
9   &       0.72    &   0.06    &   3.67    &   0.26    \\
10  &       0.75    &   0.06    &   3.68    &   0.21    \\
12  &       0.80    &   0.06    &   3.98    &   0.38    \\
14  &       0.83    &   0.07    &   3.63    &   0.27    \\
16  &       0.86    &   0.0 &   3.52    &   0.26    \\
18  &       0.89    &   0.11    &   3.39    &   0.12    \\
20  &       0.93    &   0.07    &   3.52    &   0.20    \\
30  &       1.15    &   0.08    &   3.66    &   0.44    \\
40  &       1.21    &   0.09    &   3.43    &   0.31    \\
50  &       1.28    &   0.10    &   3.51    &   0.39    \\
60  &       1.39    &   0.11    &   3.77    &   0.67    \\
70  &       1.47    &   0.11    &   4.07    &   0.69    \\
\noalign{\smallskip}\hline
  \end{tabular}

\end{table}
It is important to stress that in both networks (companies and branches) the scaling behavior for degree
distribution occurs only if we use cutoffs for links weights, compare Fig. \ref{fig:pf0} and Fig. \ref{fig:pw0a}.
It follows that such cutoffs act as filters for the noise present in the complex network topology.

\section{Entropy of network topology}
Having a probability distribution of node degrees one can calculated  a corresponding measure of network
heterogeneity. We have used the standard  formula for Gibbs entropy, i.e.

\begin{equation}\label{eq:S2}
    S=-\sum_k p(k) \ln p(k)
\end{equation}

\begin{figure}[ht]
\vskip 1cm
\centerline{\begin{tabular}{ c c}
 \epsfig{file=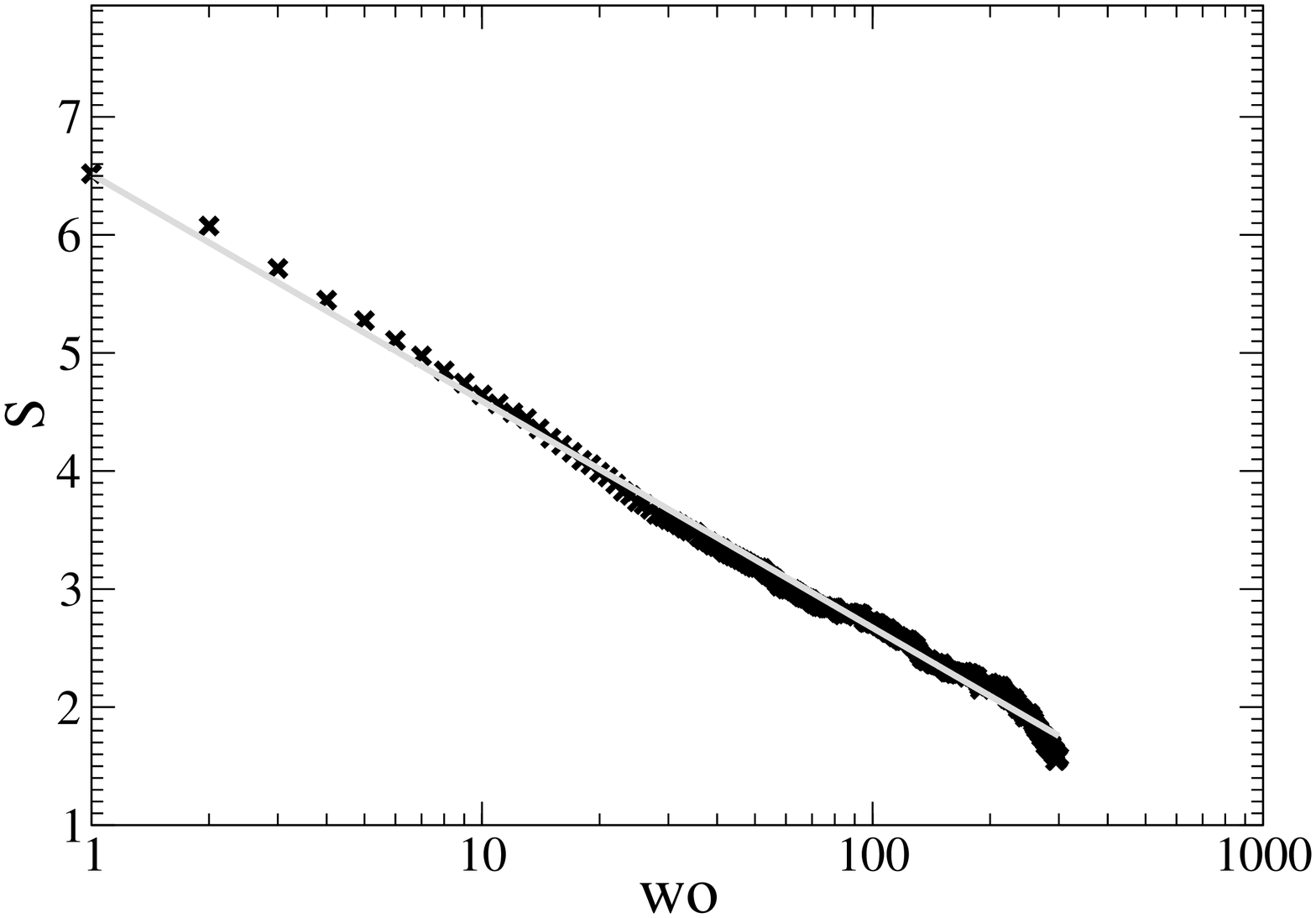,width=.52\textwidth} & \epsfig{file=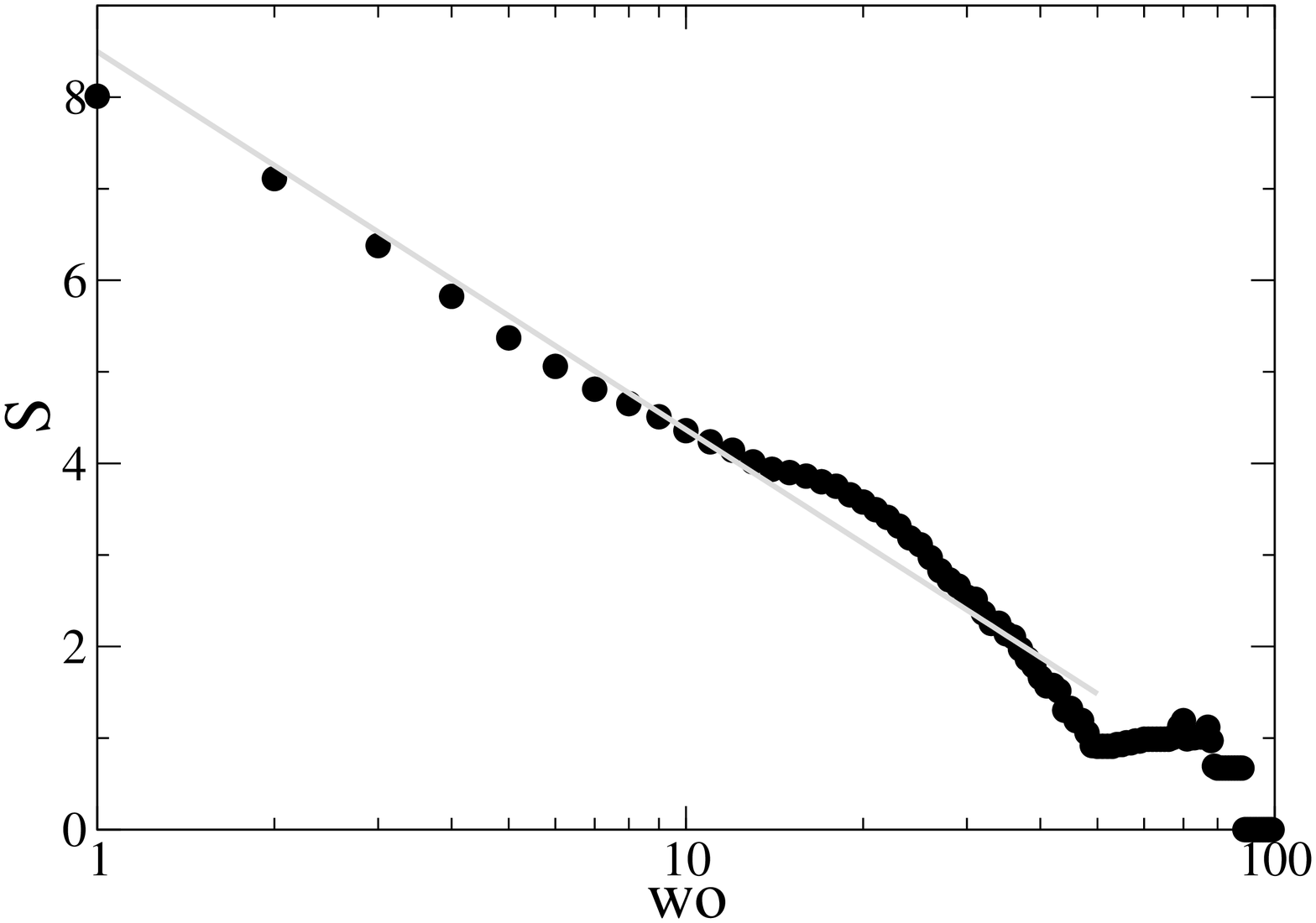,width=.52\textwidth}\\
\end{tabular}}
    \caption{Entropy dependence on cutoff parameter for branches networks on the left and for companies networks on the right.}
    \label{fig:eb}
\end{figure}

The entropy of degree distribution in branches networks decays logarithmically as a function of the cutoff value
(Fig. \ref{fig:eb})
\begin{equation}\label{eq:S3}
    S=-a \ln(w_o)+b
\end{equation}
where $a=0.834\pm 0.004$ and $b=6.51\pm0.02 $. The entropy in companies networks behaves similarly with $a=1.79\pm
0.05$ and $b=8.49\pm0.15$.

\begin{figure}[ht]
\vskip 1cm
 \centerline{\epsfig{file=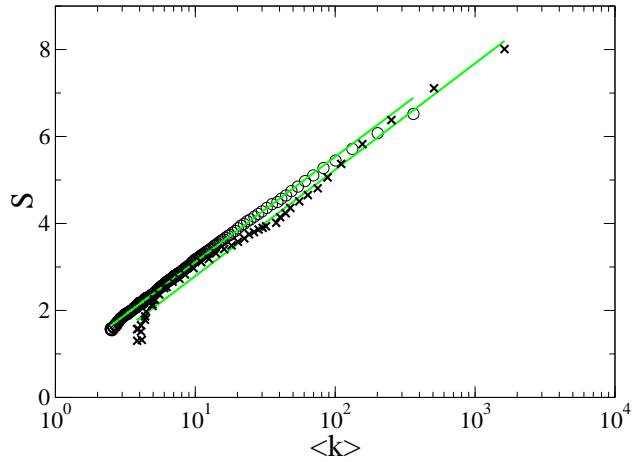,width=.7\textwidth} }
    \caption{Dependence of entropy on the average nodes degree. Circles represent branches networks and X-marks represent companies networks.}
    \label{fig:sk}
\end{figure}

The behavior has the following explanation. Diversity of node degrees is decreasing with growing weight cutoff
values $w_o$. Larger cutoffs reduce total number of links in the network what leads to a smaller range of $k$ and
thus to smaller values of $k_{max}$ and $\langle k\rangle$. The relation between $S$ and $\langle k\rangle$ is
presented at the Fig. \ref{fig:sk}, where a logarithmic scaling can be seen
\begin{equation}\label{eq:S4}
    S\sim \alpha \ln \langle k\rangle
\end{equation}

with $\alpha =1.052 \pm 0.003$ for branches networks and $\alpha =1.062 \pm 0.019$ for companies networks.

\section{Clustering coefficient}
We have analyzed a clustering coefficient dependence on node degree in branches and companies networks.
\begin{figure}[ht]
\vskip 1cm
 \centerline{\epsfig{file=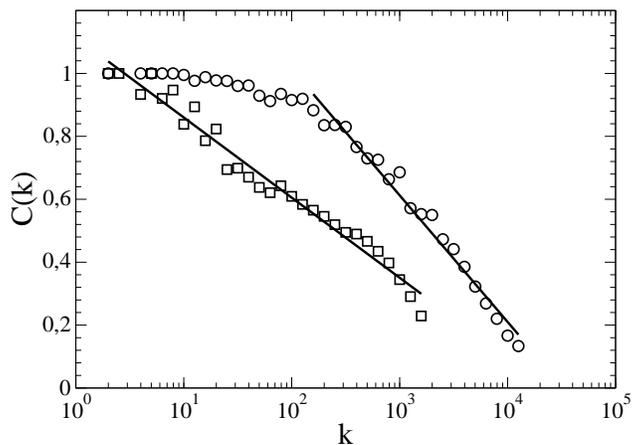,width=.7\textwidth} }

    \caption{Clustering coefficient dependence on node degree for $w_o=1$. Circles are for companies network and squares are for branch networks.}
    \label{fig:CF}
\end{figure}

In the companies network the clustering coefficient for small values of $k$ is close to one, for larger $k$ the
value of $C(k)$ exhibits logarithmic behavior
\begin{equation}\label{eq:S69}
    C\sim \beta \ln k
\end{equation}
with  $\beta_1= - 0.174\pm 0.006$. In branches networks the logarithmic behavior is present for the whole range of
$k$ with $\beta_2= - 0.111\pm 0.004$.

\section{Conclusions}
In this study, we have collected and analyzed data on companies in Poland. $48158$ medium/large firms  and $2150$
branches form a bipartite graph that allows to construct weighted networks of companies and branches.

Link weights in both networks are very heterogenous and a corresponding link weight distribution in the branches
network follows a power law. Removing links with weights smaller than a cutoff (threshold)  $w_o$ acts as a kind
of filtering for network topology. This results in recovery of a hidden scaling relations present in the network.
The degree distribution for companies networks changes with increasing $w_o$ from a nonmonotonic function with an
exponential tail (for $w_o=1$) to a power law (for $w_o>6$). For a filtered ($w_o>4$) branches network we see two
regions of scaling with different exponents and a transition point between both regimes. Entropies of degree
distributions of both networks  decay logarithmically as a function of cutoff parameter and are proportional to
the logarithm of the mean node degree.

\section{Acknowledgements}
We acknowledge a support from the EU Grant {\em Measuring and Modeling Complex Networks Across Domains} ---
MMCOMNET (Grant No. FP6-2003-NEST-Path-012999) and from Polish Ministry of Education and Science (Grant No.
13/6.PR UE/2005/7).

\bibliographystyle{ieeepes}

\end{document}